\documentclass[aps,prl,showpacs,twocolumn]{revtex4}
\usepackage{graphics}
\usepackage[dvips]{color}
\usepackage[T1]{fontenc}
\usepackage[latin1]{inputenc}
\usepackage{graphicx}
\usepackage{amssymb}
\usepackage{rotating}
\usepackage{epsfig}

\input epsf.sty

\newcommand{\ud}{{\rm d}}

\newcommand{\dir}{{Figs}}

\begin{document}

\title{Self-Consistent Field Approach for Crosslinked Copolymer Materials}
\author{Friederike Schmid}
\affiliation{Institute of Physics, JGU Mainz, D-55099 Mainz, Germany}

\begin{abstract}
A generalized self-consistent field approach for polymer networks
with fixed topology is developed. It is shown that the theory
reproduces the localization of crosslinks which is characteristic
for gels. The theory is then used to study the order-disorder
transition in regular networks of endlinked diblock copolymers.
Compared to diblock copolymer melts, the transition is shifted
towards lower values of the incompatibility parameter $\chi$ (the
Flory- Huggins parameter).  Moreover, the transition becomes
strongly first order already at the mean-field level. If stress is
applied, the transition is further shifted and finally vanishes in a
critical point.
\end{abstract}


\maketitle

Polymers are macromolecules made of a large number of structurally
identical or similar subunits (monomers), with local monomer
interactions that tend to be weak on the scale of the thermal energy
$k_B T$. In polymeric materials, molecules typically have many
interaction partners. Therefore, these systems can often be
described quite satisfactorily by mean field theories. In
particular, the self-consistent field (SCF) theory
\cite{H:75,F_book,S:98, M:02, MS:05, G_book} is a powerful mean
field approach for describing inhomogeneous polymer melts and
solutions. Originally developed as a theory for interfaces between
immiscible homopolymer phases \cite{H:75}, it has by now become a
standard tool for studying phase transitions between block copolymer
mesophases \cite{MS:94,M:95,TM:05}, the self-organization of
amphiphilic polymers in solution \cite{HS:08}, or the structure of
polymer composite materials \cite{TGMB:01,SKKF:06,KM:08}, to name
just a few examples. Numerous extensions have been proposed that
allow one to include, e.g., orientational order \cite{MF:94},
electrostatic interactions \cite{STF:04}, dynamical processes
\cite{FVM:97,MF:97,HLFB:06,ZSS:11} or the effect of fluctuations
\cite{VG:01,LKF:08,DGFS:03}.

Despite these successes, the SCF theory still suffers from severe
restrictions. Most prominently, it is limited to fluids. Polymers
are taken to have full translational freedom and, consequently, the
systems cannot sustain large shear stress or elastic deformations.
Complex fluids may respond elastically to small stress to some
extent, and this can be studied by SCF methods
\cite{BFS:05,MLKR:07}, but they necessarily yield to large stress.
In reality, however, many materials of interest such as rubber are
irreversibly crosslinked, either chemically or physically.
Crosslinking is a popular strategy for stabilizing composite
materials or polymeric nanostructures. While SCF approaches have
been devised for describing systems of reversibly crosslinked
polymeric materials \cite{MEF:10,MMF:11,LGE:12}, there exists so far
no SCF theory for irreversibly crosslinked polymer networks.

In the present paper, we propose a way to overcome this limitation.
We develop a SCF approach for irreversibly crosslinked networks with
fixed (quenched) topology.

As an application example, we then use the method to study the phase
behavior of symmetric cross-linked diblock copolymers. Specifically,
we address the question how crosslinking affects the order-disorder
transition (ODT), i.e., the transition between a disordered state
and an ordered microphase separated state. The microphase separation
in fluids of diblock copolymers has been studied intensely by SCF
methods \cite{MS:94,M:95,TM:05}. According to mean-field theory,
melts of symmetric diblock copolymers undergo a continuous
microphase separation transition to an ordered lamellar phase upon
cooling.  Fluctuations shift the transition and it becomes weakly
first order \cite{LKF:08,FH:87,BM:10}.  Whereas this is all well
understood in polymer fluids, the situation in crosslinked systems
is less clear.

Most theoretical studies of microphase separation in crosslinked
polymer blends have focussed on situations where a blend of
incompatible A and B polymers is first randomly crosslinked in a
high-temperature homogeneous state and then cooled down.  In a
seminal paper, de Gennes predicted a spinodal instability with
respect to microphase separation in such systems \cite{G:79}.
Several authors have built on this idea and investigated the
instability by experiment \cite{BB:88}, theory
\cite{BVDB:94,SSB:94,RBM:95,WZG:05,WZG:06} and simulation
\cite{LSB:00,KSE:09}. The influence of random network forces on the
structure of the ordered state was investigated by elasticity
theories \cite{PR:96,U:04}.

Here, we will consider a slightly different situation: We will study
a diblock copolymer network crosslinked in a lamellar state, asking
whether and how crosslinking stabilizes the lamellar structure. In
computer simulations, Lay {\em et al.} \cite{LJB:99} found that the
lamellar order of loosely crosslinked ordered diblock copolymer
melts disappears upon heating. We will investigate  this phenomenon.
More specifically, we will consider a regular network of endlinked
diblock copolymers. Experimentally, the synthesis of such ideal
copolymer networks is coming within reach \cite{RP:12}, and simple
phenomenological theories have been devised to predict the expected
microphase separation \cite{GVP:04,KSG:05}. A more refined SCF
theory which can predict the phase behavior of such systems based on
molecular parameters is thus clearly desirable.

To introduce the general formalism of the new SCF approach, we
consider for simplicity a regular network without dangling ends,
with connecting polymer strands made of different types of monomers
$\alpha = A,B,...$. The overall strand density is $C$. The system is
defined by (i) The topology of the network, i.e., the set of
crosslinks $l$ and linker connections $\langle jk \rangle$, (ii) the
crosslink positions $\mathbf{r}_j$ and the conformations of linker
strands, which we parameterize by space curves $\mathbf{R}_{jk}(s)$
with $s \in [0\!\!:\!\!1]$, and (iii) the sequence of monomers along
the strands, described by characteristic functions
$\eta_{jk}^{(\alpha)}(s)$, with $\eta_{jk}^{(\alpha)}(s) =1$ if the
chain is occupied by monomers of type $\alpha$ at the position $s$,
and $\eta_{jk}^{(\alpha)}(s)=0$ otherwise ( $\sum_\alpha
\eta_{jk}^{(\alpha)}(s) \equiv 1$). Strands are taken to be Gaussian
chains with polymerization index $N$ and statistical segment length $b$. The
interaction between monomers is defined by an interaction potential
$U[\{\hat{\Phi}_\alpha\}]$ which depends on the local monomer
fractions $\hat{\Phi}_{\alpha} = \frac{1}{C} \sum_{<jk>} \int_0^1
\ud s \: \delta(\mathbf{r}-\mathbf{R}_{jk}(s)) \:
\eta_{jk}^{(\alpha)}(s)$. In the following, we will mainly consider
binary AB melts with Flory-Huggins interaction $U = C \int \ud
\mathbf{r} \: \big\{ \chi N \: \hat{\Phi}_A \hat{\Phi}_B +
\frac{\kappa N}{2} ( \hat{\Phi}_A + \hat{\Phi}_B - 1)^2 \big\} $,
where $\chi$ is the Flory-Huggins parameter and $\kappa$ is related
to the inverse compressibility of the melt. Here and in the
following, the energy units are chosen such that $k_B T = 1$. The
partition function of the whole system thus reads
\begin{eqnarray}
\lefteqn{ {\cal Z} \! = \! \prod_l \!\! \int \!\! \ud
  \mathbf{r}_l \!\!\! \prod_{<jk>}  \!\!  \int {\cal D} \mathbf{R}_{jk} \:
    \delta(\mathbf{r}_j-\mathbf{R}_{jk}(0)) \: \:
    \delta(\mathbf{r}_k-\mathbf{R}_{jk}(1)) } 
\nonumber \\ &&
  \qquad \times 
    \exp\Big(\!\!-\!\! \sum_{<mn>} \frac{1}{R_g^2} \int_0^1 \!\!\! \ud s \: (\frac{\ud
\mathbf{R}_{mn}}{\ud s})^2 + U[\{\hat{\Phi}_\alpha \}] \Big),
\hspace*{0.5cm}
\label{eq:partition_function}
\end{eqnarray}
where $R_g = b \sqrt{N/6}$ is the radius of gyration of one strand.
After integrating out the crosslink degrees of freedom, we recover
the partition function of the classical Deam-Edwards theory for 
networks \cite{DE:76, GZ:93, GCZ:96, PRa:96}. In the present approach, 
it will prove convenient to keep the crosslink degrees of freedom 
explicitly.

We first derive a mean-field approximation for the contributions of
monomer interactions to the free energy in a standard
field-theoretic SCF manner \cite{S:98}: The chain conformations are
decoupled by inserting identity operators $1 \propto \int_{\infty}
\!\! {\cal D} \Phi_\alpha \int_{i \infty} \!\! {\cal D} W_\alpha \:
e^{C \int \ud \mathbf{r} \: W_\alpha (\Phi_\alpha -
\hat{\Phi}_\alpha)}$, where $\int_{\infty} {\cal D} \Phi_\alpha$ and
$\int_{i \infty} {\cal D} W_\alpha$ denote functional integrals over
fluctuating fields $\Phi_\alpha(\mathbf{r})$ and
$W_\alpha(\mathbf{r})$. This allows us to rewrite the partition
function in the form \mbox{${\cal Z} \propto \prod_\alpha
\int_{\infty} \!\! {\cal D} \Phi_\alpha \int_{i \infty} \!\! {\cal
D} W_\alpha \: \prod_l \int \ud \mathbf{r}_l e^{- {\cal
F}[\{\Phi_\alpha, W_\alpha, \mathbf{r}_l\}]}$} with
\begin{equation}
\label{eq:energy_scf}
{\cal F} = U[\{\Phi_\alpha\}]
- C \sum_\alpha \int \!\! \ud \mathbf{r} \: \Phi_\alpha W_\alpha
- \sum_{<jk>} \ln {\cal Q}_{jk}(\mathbf{r}_j,\mathbf{r}_k).
\end{equation}
Here the ${\cal Q}_{jk}(\mathbf{r},\mathbf{r'})$ are single-chain
partition functions of the linker strands $\mathbf{R}_{jk}(s)$ in
the self-consistent field $W_\alpha$, subject to the constraint that
the ends are located at the crosslink positions $\mathbf{r}$ and
$\mathbf{r}'$:
\begin{equation}
\label{eq:qjk}
{\cal Q}_{jk}(\mathbf{r},\mathbf{r}') =
\int {\cal D} \mathbf{R}
\Big|_{{\mathbf{R}(0)=\mathbf{r}} \atop {\mathbf{R}(1)=\mathbf{r}'}}
e^{- \int_0^1 \!\! \ud s \: \eta_{jk}^{(\alpha)}(s) W_\alpha(\mathbf{R}(s))}.
\end{equation}

The SCF approximation consists in replacing the fluctuating
field integral by its saddle point, i.e., the free energy is
approximated by the minimum of ${\cal F}$ in Eq.~(\ref{eq:energy_scf})
with respect to the fields $\Phi_\alpha$ and $W_\alpha$.
One obtains the set of self-consistent SCF equations
$C W_\alpha(\mathbf{r}) = \delta U/\delta \Phi_\alpha(\mathbf{r})$ and
$\Phi_\alpha(\mathbf{r}) = \sum_{<jk>}
\delta \ln({\cal Q}_{jk}(\mathbf{r}_j,\mathbf{r}_k))
/\delta W_\alpha(\mathbf{r})$. Efficient strategies to evaluate
${\cal Q}_{jk}$ and $\Phi_\alpha$ can be found in the
literature \cite{MS:05}.

The remaining task is to find a corresponding mean-field approximation for the
integration over the {\em crosslink} degrees of freedom. This is done
following an analogous field-theoretic procedure. We define local crosslink
distribution operators $\hat{p}_j(\mathbf{r}) =
\delta(\mathbf{r}-\mathbf{r}_j)$ and insert identities
$1 \propto \int_{\infty} \!\! {\cal D} p_j
\int_{i \infty} \!\! {\cal D} h_j \:
e^{\int \ud \mathbf{r} \: h_j (p_j - \hat{p}_j)}$.
The partition function is thus rewritten as
${\cal Z} \propto \prod_j \int_{\infty} {\cal D} p_j
\int_{i \infty} {\cal D} h_j \: e^{- F[p_j, h_j]}$
with the new free energy functional
\begin{eqnarray}
F & = & U[\{\Phi_\alpha\}]
- C \sum_\alpha \int \!\! \ud \mathbf{r} \: \Phi_\alpha W_\alpha
\nonumber \\&&
- \sum_{<ij>}
\int \! \int \! \ud \mathbf{r} \: \ud \mathbf{r}' \:
p_j(\mathbf{r}) \: p_k(\mathbf{r}') \: \ln {\cal Q}_{jk}(\mathbf{r},\mathbf{r}')
\nonumber \\ &&
- \sum_l \int \! \ud \mathbf{r} \: h_l(\mathbf{r}) \: p_l(\mathbf{r})
- \sum_l \ln({\cal N}_l)
\label{eq:energy_bw}
\end{eqnarray}
with ${\cal N}_l = \int \ud \mathbf{r} \: e^{- h_l(\mathbf{r})}$.
The mean-field approximation then again consists in carrying out a
saddle point integration, {\em i.e.}, minimizing $F$ in Eq.
(\ref{eq:energy_bw}) with respect to $p_j$ and $h_j$. This results
in the mean-field equations
\begin{eqnarray}
\label{eq:bw1}
h_j(\mathbf{r}) &=& - \sum_{\{k\}_j}
\int \ud \mathbf{r}' \: p_k(\mathbf{r}') \:
\ln ({\cal Q}_{jk}(\mathbf{r},\mathbf{r}'))
\\
p_j(\mathbf{r}) &=& {\cal N}_j^{-1} e^{- h_j(\mathbf{r})},
\label{eq:bw2}
\end{eqnarray}
where the sum $\sum_{\{k\}_j}$ runs over the crosslinks $k$
that are directly linked with $j$.

As a side note, we remark that the free energy expression in Eq.\
(\ref{eq:energy_scf}) has the structure of a Hamiltonian for a
lattice model with sites $j$, continuous degrees of freedom
$\mathbf{r}_j$, and ``interactions'' $\ln {\cal Q}_{jk}$. Thus, mean
field methods developed for lattice models can be applied in the
present problem as well. The approximation derived above is
equivalent to the popular Bragg-Williams approximation \cite{P_book}
in a version for continuous degrees of freedom. It yields
the same equations, (\ref{eq:energy_bw}-\ref{eq:bw2}). Other more
sophisticated approximations such as the Bethe approximation
\cite{P_book} can be adopted as well. More details are found in
the supplemental information. Unless stated otherwise, we will
use Eqs.\ (\ref{eq:energy_bw}) -- (\ref{eq:bw2}) here, i.e., 
the Bragg-Williams approximation.

In the absence of any monomer interactions (phantom networks), the
mean-field equations (\ref{eq:bw1}) and (\ref{eq:bw2}) can be solved
analytically. For regular networks with equivalent crosslinks, the
crosslinks are centered about their respective mean
position $\mathbf{R}_j$ with a Gaussian distribution,
\begin{equation}
\label{eq:bw_ideal}
p_j(\mathbf{r}) = R_g^{-d}  \sqrt{\frac{f}{4 \pi}}^d
\exp\big({-\frac{f}{4 R_g ^2} \:  (\mathbf{r}-\mathbf{R}_j)^2}\big),
\end{equation}
where $f$ is the functionality of the crosslink and $d$ the spatial dimension.
Thus the theory reproduces the phenomenon of crosslink localization on a scale
of $R_g$, in agreement with theories of the sol-gel transition in networks
\cite{GZ:93,GCZ:96,PRa:96}.

\begin{figure}[tb]
\includegraphics[scale=0.5]{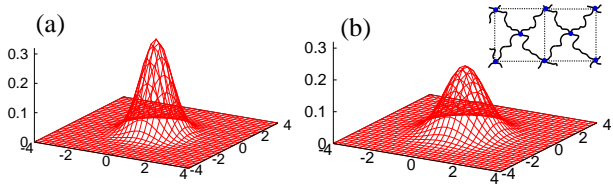}
\vspace*{-0.3cm} \caption{\label{fig:distributions} Distribution of
individual crosslinks about their average position in a 2D melt of
almost incompressible endlinked homopolymers ($\kappa N = 100$) at
polymer density $C R_g^2 = 4$, calculated within the Bragg-Williams
approximation (a), and the Bethe approximation (b). The network has
the topology of a square lattice (see inset in b). The distribution
in (a) is indistinguishable from Eq.\ (\protect\ref{eq:bw_ideal})
with $f=4$ and $d=2$.}
\end{figure}

Interacting networks swell due to the excluded volume interactions.
As a result, the mean crosslink positions $\mathbf{R}_j$ move apart.
In incompressible materials, however, the {\em shape} of the
crosslink distribution is barely affected by the swelling. Fig.\
\ref{fig:distributions} shows numerical results for an interacting
network with square topology in two dimensions (2D) \cite{footnote} in 
the Bragg-William and the Bethe approximation (equations are given
in the supplemental information). The Bragg-William result
is practically indistinguishable from the prediction of Eq.\
(\ref{eq:bw_ideal}).

\begin{figure}[tb]
\includegraphics[scale=0.35]{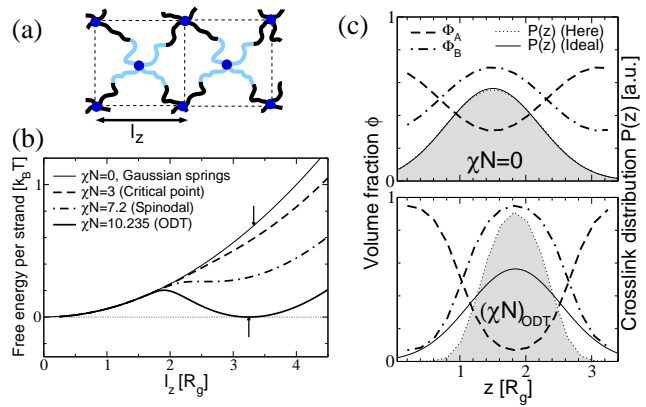}
\vspace*{-0.3cm} \caption{\label{fig:2D_energy} Properties of a
stretched AB diblock copolymer network in the 2D square topology
sketched in (a) at $\kappa N = 100$ and strand density $C R_g^2 \to
\infty$. The case $\chi N=0$ also describes homopolymer networks.
(b) Free energy per strand vs.\ imposed cell length $l_z$ for
different values of $\chi N$ as indicated. Arrows mark $l_z=3.245
R_g$. (c) Composition profile ($\Phi_{A,B}$: dashed and dot-dashed
lines) and crosslink distribution profile ($P$: dotted and shaded
curve) across a lamella at imposed cell length $l_z = 3.245 R_g$ for
$\chi N=0$ (top), and $\chi N = 10.235$ (bottom). For comparison,
thin solid line shows crosslink distribution for non-stretched ideal
(phantom) chains (Eq.\ (\protect\ref{eq:bw_ideal})). }
\end{figure}

Likewise, the crosslink distribution does not change if the network
is deformed elastically. Fig.\ \ref{fig:2D_energy}(c)(top panel)
shows the crosslink distribution for an (almost) incompressible
homopolymer network that has been stretched considerably in one
direction. It can be described almost perfectly by Eq.\
(\ref{eq:bw_ideal}). Specifically, we consider the network ``unit
cell'' sketched in Fig.\ \ref{fig:2D_energy} with side length $l_z$.
In the mechanically relaxed state, $l_z$ takes the equilibrium value
$l_z = l^* = 2/\sqrt{C}$. If the cells are stretched to $l_z \gg
l^*$, the free energy per strand as a function of $l_z$ rises
quadratically according to $F/n_{\mbox{\tiny Strands}} \approx
l_z^2/(16 R_g^2)$ (see Fig.\ \ref{fig:2D_energy}(b), case $\chi N=0$),
which corresponds to the behavior of phantom networks of Gaussian
chains. Incompressible homopolymer networks behave very much like
phantom networks. Deviations only start to set in for strongly
swollen networks with strand densities $C$ below a crossover value
$C^* = R_g^{-d}$ (with the spatial dimension $d$), which is the
network equivalent of the overlap concentration in polymer solutions
\cite{degennes_book}.

This changes in networks of AB copolymers. We consider the same (2D)
square network as above\cite{footnote}, now made of AB diblock copolymers
which are endlinked such that A ends join A and B ends join B (Fig.\
\ref{fig:2D_energy}(a). By stretching the network in the $z$
direction, an average distance $l_z/2$ is imposed between
neighboring A and B crosslinks, which induces a separation between A
rich and B rich regions (Fig. \ref{fig:2D_energy}(c). At $\chi N=0$,
the free energy rises monotonically as a function of $l_z$ as
discussed above (homopolymer case). As $\chi N$ is increased, a
minimum at nonzero $l_z$ develops and gradually deepens (Fig.\
\ref{fig:2D_energy}(b). This minimum dominates beyond $\chi N =
10.235$, and the network undergoes an order-disorder transition
(ODT) from a disordered state (with $l_z =l^*\ll R_g$) to a
microphase separated lamellar state with lamellar spacing $l_z=3.245
R_g$. Compared to the noninteracting stretched structure at same
$l_z$, the stable lamellar state at the ODT is characterized by
higher segregation, and by a narrower, more localized crosslink
distribution.

\begin{figure}[t]
\includegraphics[scale=0.4]{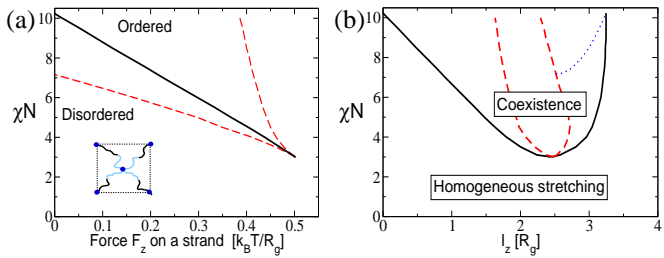}
\vspace*{-0.3cm} \caption{\label{fig:2D_phases} Phase diagram for
the system of Fig.\ \protect\ref{fig:2D_energy} in the plane of
$\chi N$ vs. (a) the average force acting on a strand $F_z$, (b) the
imposed cell length $l_z$. Solid lines: coexistence. Dashed lines:
spinodals. Dotted line in (b): lamellar distance $l_z$ of
(metastable or stable) ordered state in the stress free system. }
\end{figure}

Comparing the ODT in the network to the ODT in copolymer melts, one
notices striking dissimilarities. First, the transition is strongly
first order already at the mean-field level. Second, stretching the
system stabilizes the ordered phase. Free energy curves such as
those shown in Fig.\ \ref{fig:2D_energy}(b) can be used to construct
a phase diagram as a function of the average force $F_z$ acting on a
strand. The result is shown in Fig.\ \ref{fig:2D_phases}. As the
system is stretched, the first order transition gradually moves to
lower $\chi N$ until it finally ends in a critical point at $(\chi
N)_c=3$. The critical point in the network is thus found at much
lower $\chi N$ than the critical point in the melt ($(\chi
N)_c=10.425$). However, the ODT in the stress-free system (at
$F_z=0$), is only slightly shifted compared to the melt.

\begin{figure}[b]
\includegraphics[scale=0.4]{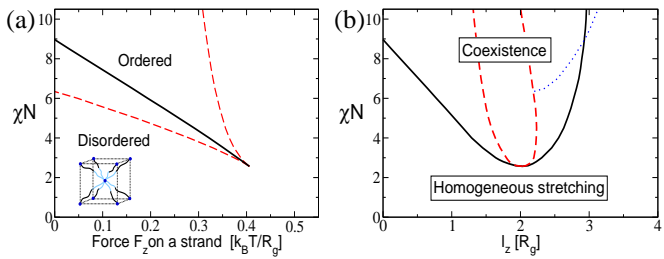}
\vspace*{-0.3cm} \caption{\label{fig:3D_phases} Same as Fig.\
\protect\ref{fig:2D_phases} for a three-dimensional network with a
bcc topology (see cartoon inset in (a)). }
\end{figure}

The exact value of the ODT depends on the network topology and on
the spatial dimension. For comparison, we have also calculated the
phase behavior for a three dimensional network with body-centered
cubic (bcc) topology (Fig.\ \ref{fig:3D_phases}). The resulting
phase diagram is similar to the two-dimensional one. The shift of
the ODT is more pronounced, but still not spectacular. Thus we find
that weak crosslinking does not stabilize lamellar order
efficiently, in agreement with the simulation results of Lay {\em et
al.} \cite{LJB:99}.

Finally, we discuss the elastic behavior of the network. We focus on
the three-dimensional case, and calculate Young's moduli for our bcc
topology. In the isotropic disordered state, the modulus is given by
$B_i=1.5 C^{1/3} /R_g^2$, indicating that the network becomes stiff
at high strand density $C$. In the ordered state, the modulus
depends on the direction of applied stress. The elastic response to
stretching in the direction normal to the lamellae is stiffer than
in the isotropic case, $B_n= 2.12 C$. In contrast, the system is
soft in the parallel directions, with a modulus that does not depend
on $C$ at all, $B_{\parallel} = 1.36/R_g^2$. Hence, the elastic
penalty for stretching lamellae in plane is found to be much smaller
than the penalty for stretching them in the normal direction. This
is compatible with a prediction of Panyukov and Rubinstein
\cite{PR:96} based on a phenomenological model for networks. On
applying stress to a microphase separated system with initially
randomly oriented lamellae, one thus expects the lamellae to be
kinetically driven towards aligning parallel to the applied stress.
This was indeed observed experimentally \cite{SAOO:01}. The true
state of lowest free energy, however, is one where the lamellae are
oriented normal to the stress, such that the (elongated) copolymer
strands are aligned parallel to the applied force. Lamellae with
parallel orientation can only reach this state by copolymer
reordering, which involves crossing an energy barrier. Hence the
parallel orientation might be stabilized kinetically in many cases,
even though the true equilibrium state is one with normal
orientation.

To summarize, we have presented a way to extend the SCF theory to
quenched polymer networks. It opens up a wide range of new
applications for the SCF approach, such as the study of chemically
or physically crosslinked systems, or of liquid crystalline
elastomers (using suitable extensions of the SCF theory
\cite{MF:94}). We have used the approach to study the ODT in
networks of diblock copolymers, and the results were compatible with
available simulations, experiments, and phenomenological theories.

In the present paper, the theory was developed for regular networks.
The next step in future work will be to include disorder, and
disorder averages, e.g., within a CPA (Coherent Potential
Approximation) type approach. In this context, it will be interesting
to establish the connection to classical field theories of networks
based on the Deam-Edwards approach \cite{DE:76, GZ:93, GCZ:96},
which focus on the process of generating topological disorder
during crosslinking. Another important issue is the effect
of entanglements. The present theory ignores topological
interactions by construction (since the monomers have no hard-core
interactions). However, the effect of entanglements may be similar to
that of physical crosslinks, e.g., in interpenetrating networks
\cite{BF:84,S:92, SB:93}, and it might be possible to treat them as
effective crosslinks in some cases.

The present theory can also be used as a starting point for simpler
Landau-type expansions. For example, the derivation of a RPA 
(Random Phase Approximation) theory \cite{Leibler} for networks 
might give additional insight into the nature of the phase transitions 
in the system.

This work was started during a visit to the Materials Research Lab
at UCSB Santa Barbara (USA). Inspiring discussions with G.
Fredrickson and his group are gratefully acknowledged.

\section*{Supplemental information}

\noindent
In the following, we give and motivate the mean-field equations for polymer crosslink
distributions in the Bethe approximation, which were used to obtain Fig.\ 1\ (b). To put
the Bethe approximation into context, we begin with giving an alternative derivation
of the Bragg-Williams equations, Eqs.\ (4-6), which does not rely on field theory,
but adopts a more traditional approach \cite{p_book}. 

The starting point is Eq.\ (2), which defines an ''effective Hamiltonian''
of the form
\begin{equation}
{\cal F} = {\cal F}_0 
  - \sum_{<jk>} \ln {\cal Q}_{jk}(\mathbf{r}_j,\mathbf{r}_k),
\end{equation}
where the  sum $<\!jk\!>$ runs over connected neighbors and 
${\cal F}_0$ does not depend explicitly on the crosslink positions $\mathbf{r}_j$.
(It does of course depend implicitly on the crosslink positions, once the
self-consistent solution for the densities $\Phi$ and fields $W$ has been 
inserted, but here we may take $\Phi$ and $W$ to be fixed).

The free energy of the system is given by
\begin{equation}
\label{eq:free}
F = \langle {\cal F} \rangle - S,
\end{equation}
with the average energy $\langle {\cal F} \rangle$  and the entropy $S$. 
As before, the temperature has been set to one. Our task is to evaluate and
minimize this expression with respect to the global distribution function
of the $M$ crosslinks in the system, $P(\mathbf{r}_1,\cdots,\mathbf{r}_M)$.

In the Bragg-Williams approximation, one assumes that the global distribution 
function factorizes,
$P(\mathbf{r}_1,\cdots,\mathbf{r}_m) = \prod_l p_l(\mathbf{r}_l)$.
Inserting this gives the average energy
\begin{equation}
\langle {\cal F} \rangle_{\mbox{\tiny BW}} \approx
{\cal F}_0 
- \sum_{<jk>}
\int \! \int \! \ud \mathbf{r} \: \ud \mathbf{r}' \:
p_j(\mathbf{r}) \: p_k(\mathbf{r}') \: \ln {\cal Q}_{jk}(\mathbf{r},\mathbf{r}')
\end{equation}
and the entropy of the crosslink distribution
\begin{equation}
S_{\mbox{\tiny BW}} \approx
-\sum_l \int \ud \mathbf{r} \: p_l(\mathbf{r}) \: \ln(p_l(\mathbf{r})).
\end{equation}
The minimization of Eq.\ (\ref{eq:free}) with respect to the functions $p_l(\mathbf{r})$
under the constraint $\int \ud \mathbf{r} \; p_l (\mathbf{r}) = 1$ for all $l$
results in Eqs.\ (5) and (6), and the free energy (4). 

Whereas all correlations between crosslinks are neglected in the
Bragg-Williams approximation, the Bethe approximation accounts for
correlations between crosslinks that are directly connected 
(i.e., that have direct interactions with each other). Hence
the factorization approximation is dropped for these pairs, and
the global distribution function is approximated by 
\begin{equation}
\label{eq:bethe}
P(\mathbf{r}_1,\cdots,\mathbf{r}_M) = 
 \prod_{<jk>} p_{jk}^{(2)}(\mathbf{r}_j, \mathbf{r}_k) \Big/
 \prod_{l} [p_{l}(\mathbf{r}_l)]^{c_l-1},
\end{equation}
where $p_{jk}^{(2)}$ are the pair distribution functions, and 
$c_l$ is the connectivity of crosslink $l$. We note that
$p^{(2)}_{mn}(\mathbf{r}_m, \mathbf{r}_n)/p_n(\mathbf{r}_n)$ is the
conditional probability of finding the crosslink $m$ at position 
$\mathbf{r}_m$, if crosslink $n$ is at position $\mathbf{r}_n$.
Using this relation, one can easily check that 
the approximation (\ref{eq:bethe}) fulfills
$
\prod_{l \ne (j,k)}\int \ud\mathbf{r}_l P(\mathbf{r}_1,\cdots \mathbf{r}_m)
= p_{jk}^{(2)}(\mathbf{r}_j,\mathbf{r}_k).
$
The approximation (\ref{eq:bethe}) results in the expression \cite{p_book}
\begin{equation}
\langle {\cal F} \rangle_{\mbox{\tiny Bethe}} \approx
{\cal F}_0 
- \sum_{<jk>}
\int \! \int \! \ud \mathbf{r} \: \ud \mathbf{r}' \:
p_{jk}^{(2)}(\mathbf{r},\mathbf{r}') \: \ln {\cal Q}_{jk}(\mathbf{r},\mathbf{r}')
\end{equation}
for the energy, and
\begin{eqnarray}
\nonumber
S_{\mbox{\tiny Bethe}} \approx
&-& \sum_{<jk>}
\int \! \int \! \ud \mathbf{r} \: \ud \mathbf{r}' \:
p_{jk}^{(2)}(\mathbf{r},\mathbf{r}') \: \ln (p_{jk}^{(2)}(\mathbf{r},\mathbf{r}') ) 
\\ && 
+\sum_l (c_l-1) \int \ud \mathbf{r} \: p_l(\mathbf{r}) \: \ln(p_j(\mathbf{r})).
\end{eqnarray}
for the entropy.
After minimizing again Eq.\ (\ref{eq:free}) with respect to $p_{jk}^{(2)}$ and $p_l$,
subject to the constraints
$\int \ud \mathbf{r}  \: p_{jk}^{(2)}(\mathbf{r},\mathbf{r}')= p_k(\mathbf{r}')$,
$\int \ud \mathbf{r}' \: p_{jk}^{(2)}(\mathbf{r},\mathbf{r}')= p_j(\mathbf{r})$,
and $\int \ud \mathbf{r} \; p_l (\mathbf{r}) = 1$, we finally obtain the Bethe 
equations
\begin{equation}
p_{jk}^{(2)}(\mathbf{r},\mathbf{r}') =
{\cal M}_{jk}(\mathbf{r}) {\cal Q}_{jk}(\mathbf{r},\mathbf{r}') 
\overline{\cal M}_{jk}(\mathbf{r}')
\end{equation}
\begin{equation}
p_l(\mathbf{r}) = 
{\cal N}_l \prod_{\{j\}_l} {\cal M}_{jl}(\mathbf{r})^{1/(c_l-1)} \:
 \prod_{\{j'\}_l} \overline{\cal M}_{j'l}(\mathbf{r})^{1/(c_l-1)}.
\end{equation}
Here the products $\prod_{\{j\}_l}$ and $\prod_{\{j'\}_l}$ run
over the connected neighbors of the crosslink $l$, and the functions
${\cal M}_{jk}(\mathbf{r}), \overline{\cal M}_{jk}(\mathbf{r})$, as well as the 
constants ${\cal N}_l$ must be determined self-consistently such that the 
constraints are satisfied. The resulting free energy in Bethe approximation
takes the simple form
\begin{equation}
F_{\mbox{\tiny Bethe}} = {\cal F}_0 - \sum_l (c_l - 1) {\cal N}_l.
\end{equation}


\begin{thebibliography}{66}

\bibitem{H:75} E. Helfand, J. Chem. Phys. {\bf 62}, 999 (1975).
\bibitem{F_book} G. Fleer, M. Cohen Stuart, J. Scheutjens, T. Cosgrove, B. Vincent,
{\em Polymers at interfaces} (Kluwer Academic Publ., Reading, Massachusetts, 1993).
\bibitem{S:98} F. Schmid, J. Phys.: Condens. Matter {\bf 10}, 8105 (1998).
\bibitem{M:02} M.W. Matsen, J. Phys.: Cond. Matter {\bf 14}, R21 (2002).
\bibitem{MS:05} M. M\"uller and F. Schmid, Adv. Polym. Sci. {\bf 185}, 1 (2005).
\bibitem{G_book} G.H. Fredrickson, {\em The equilibrium theory of inhomogeneous polymers}
(Oxford University Press, Oxford, UK, 2006).
\bibitem{MS:94} M.W. Matsen, M. Schick, Phys. Rev. Lett. {\bf 72}, 2660 (1994).
\bibitem{M:95} M.W. Matsen, Macromolecules {\bf 28}, 5765 (1995).
\bibitem{TM:05} C.A. Tyler, D.C. Morse, Phys. Rev. Lett. {\bf 94}, 208302 (2005).
\bibitem{HS:08} X.H. He, F. Schmid, Phys. Rev. Lett. {\bf 100}, 137802 (2008).
\bibitem{TGMB:01} R.B. Thompson, W. Ginzburg, M.W. Matsen, A.C. Balazs,
  Science {\bf 292}, 2469 (2001).
\bibitem{SKKF:06} S.W. Sides, B.J. Kim, E.J. Kramer, and G.H. Fredrickson,
  Phys. Rev. Lett. {\bf 96}, 250601 (2006).
\bibitem{KM:08} J.U. Kim and M.W. Matsen,
  Macromolecules {\bf 41}, 4435 (2008). 
\bibitem{MF:94} D.C. Morse, G.H. Fredrickson,
  Phys. Rev. Lett. {\bf 73}, 3235 (1994).
\bibitem{STF:04} Q. Wang, T. Taniguchi, G.H. Fredrickson,
  J. Phys. Chem. B {\bf 108}, 6733 (2004).
\bibitem{FVM:97} J.G.E.M. Fraaije, B.A.C. van Vlimmeren,
N.M. Maurits, M. Postma, O.A. Evers, C. Hoffmann, P. Altevogt, G. GoldbeckWood,
J. Chem. Phys. {\bf 106}, 4260 (1997).
\bibitem{MF:97} N.M. Maurits, J.G.E.M. Fraaije, J. Chem. Phys. {\bf 107}, 5879 (1997).
\bibitem{HLFB:06} D.M. Hall, T. Lookman, G.H. Fredrickson, S. Banerjee,
  Phys. Rev. Lett. {\bf 97}, 114501 (2006).
\bibitem{ZSS:11} L. Zhang, A. Sevink, F. Schmid, Macromolecules{\bf 44}, 9434 (2011).
\bibitem{VG:01} V. Ganesan, G.H. Fredrickson, Europhys. Lett. {\bf 55}, 814 (2001).
\bibitem{DGFS:03} D. D\"uchs, V. Ganesan, G.H. Fredrickson, and F. Schmid,
   Macromolecules {\bf 36}, 9237 (2003).
\bibitem{LKF:08} E.M. Lennon, K. Katsov, and G.H. Fredrickson, Phys. Rev. Lett. {\bf 101}, 138302 (2008).
\bibitem{BFS:05} J.L. Barrat, G.H. Fredrickson, S.W. Sides,
  J. Phys. Chem. B {\bf 109}, 6694 (2005). 
\bibitem{MLKR:07} P. Maniadis, T. Lookman, E.M. Kober, K.O. Rasmussen,
  Phys. Rev. Lett. {\bf 99}, 048302 (2007).
\bibitem{MEF:10} A. Mohan, R. Elliot, G.H. Fredrickson, J. Chem. Phys. {\bf 133}, 174903 (2010).
\bibitem{MMF:11} Z. Mester, A. Mohan, G.H. Fredrickson, Macromolecules {\bf 44}, 9411 (2011).
\bibitem{LGE:12} D. Li, T. Gruhn, H. Emmerich, J. Chem. Phys. {\bf 137}, 024906 (2012).
\bibitem{FH:87} G.H. Fredrickson, E. Helfand, J. Chem. Phys. {\bf 87}, 697 (1987).
\bibitem{BM:10} T.M. Beardsley, M.W. Matsen, Eur. Phys. J. E {\bf 32}, 255 (2010).

\bibitem{footnote} We emphasize that we consider a true 2D system here (in an
$(x,z)$ plane), not a tethered membrane in 3D space.
\bibitem{G:79} P.G. de Gennes, J. Physique Lettres {\bf 40}, 69 (1979).
\bibitem{BB:88} R.M. Briber, B.J. Bauer, Macromolecules {\bf 21}, 3296 (1988).
\bibitem{BVDB:94} M. Benmouna, T.A. Vilgis, M. Daoud, M. Benhamou,
Macromolecules {\bf 27}, 1172 (1994).
\bibitem{SSB:94} S. Stepanow, M. Schulz, K. Binder, J. Physique II {\bf 4}, 819 (1994)
\bibitem{RBM:95} D.J. Read, M.G. Brereton, T.C.B. McLeish,
 J. Physique II {\bf 5}, 1679 (1995).
\bibitem{WZG:05} C. Wald, A. Zippelius, P.M. Goldbart,
  Europhys. Lett. {\bf 70}, 843 (2005).
\bibitem{WZG:06} C. Wald, P.M. Goldbart, A. Zippelius,
  J. Chem. Phys. {\bf 124}, 214905 (2006).
\bibitem{LSB:00} S. Lay, J.-U. Sommer, A. Blumen, J. Chem. Phys. {\bf 113}, 11355 (2000).
\bibitem{KSE:09} A.V. Klopper, C. Svaneborg, R. Everaers, Eur. Phys. J. E {\bf 28}, 89 (2009).
\bibitem{PR:96} S. Panyukov, M. Rubinstein, Macromolecules{\bf 29}, 8220 (1996).
\bibitem{U:04} N. Uchida, J. Phys.: Cond. Matter {\bf 16}, L21 (2004).
\bibitem{LJB:99} S. Lay, J.-U. Sommer, A. Blumen, J. Chem. Phys. {\bf 110}, 12173 (1999).
\bibitem{RP:12} M. Rikkou-Kalourkoti, C.S. Patrickios, Macromolecules {\bf 45}, 7890 (2012).
\bibitem{GVP:04} T.K. Georgiou, M. Vamvakaki, C.S. Patrickios, Polymer {\bf 45}, 7341 (2004).
\bibitem{KSG:05} M. Karbarz, Z. Stojek, C.S. Patrickios, Polymer {\bf 46}, 7456 (2005).
\bibitem{DE:76} R.T. Deam, S.F. Edwards, Phys. Trans. Royal Soc. {\bf 280}, 317 (1976).
\bibitem{GZ:93} P.M. Goldbart, A. Zippelius, Phys. Rev. Lett. {\bf 71}, 2256 (1993).
\bibitem{GCZ:96} P.M. Goldbart, H.E. Castillo, A. Zippelius, Adv. Physics {\bf 45}, 393 (1996).
\bibitem{PRa:96} S. Panyukov, Y. Rabin, Physics Reports {\bf 269}, 1 (1996).
\bibitem{P_book} M. Plischke, B. Bergersen, {\em Equilibrium Statistical Physics},
  (World Scientific, Singapore, 2005).
\bibitem{degennes_book} P.G. de Gennes, {\em Scaling concepts in polymer physics}
(Cornell Univ. Press, Cornell, 1979).
\bibitem{SAOO:01} S. Sakurai, S. Aida, S. Okamoto, T. Ono, K. Imarzumi, S. Nomura,
   Macromolecules {\bf 34}, 3672 (2001).
\bibitem{BF:84} K. Binder, H.L. Frisch, J. Chem. Phys. {\bf 81}, 2126 (1984).
\bibitem{S:92} M. Schulz, J. Chem. Phys. {\bf 97}, 5631 (1992).
\bibitem{SB:93} M. Schulz, K. Binder, J. Chem. Phys. {\bf 98}, 655 (1993).
\bibitem{Leibler} L. Leibler, Macromolecules {\bf 13}, 1602 (1980).
\end{thebibliography}

\begin{thebibliography}{66}
\bibitem{p_book} M. Plischke, B. Bergersen, {\em Equilibrium Statistical Physics},
  (World Scientific, Singapore, 2005).
\end{thebibliography}
\end{document}